\documentclass[aps,preprint]{revtex4}
\newcommand{\be}{\begin{equation}}
\newcommand{\ee}{\end{equation}}
\newcommand{\bea}{\begin{eqnarray}}
\newcommand{\eea}{\end{eqnarray}}

\begin{document}
\draft

\title{Spectral properties of quasi-one-dimensional conductors with a finite transverse band dispersion}
\author{\v{Z} Bona\v{c}i\'{c} Lo\v{s}i\'{c}$^1$,  A Bjeli\v{s}$^2$ and  P  \v{Z}upanovi\'{c}$^1$}
 \address{$^1$ Department of Physics, Faculty of Natural Sciences, Mathematics and Kinesiology,
University of Split, Teslina 12, 21000 Split, Croatia}
\email{agicz@pmfst.hr}
 \address{$^2$Department of Physics, Faculty of Science, University
 of Zagreb, POB 162, 10001 Zagreb, Croatia}
\email{bjelis@phy.hr}

\begin{abstract}
We determine the one-particle spectral function and the corresponding derived quantities for the conducting chain lattice with the
finite inter-chain hopping $t_\perp$ and the three-dimensional long-range Coulomb electron-electron interaction. The standard $G_{0}W_{0}$
approximation is used. It is shown that, due to the optical character of the anisotropic plasmon dispersion caused by the finite $t_\perp$, the
low energy quasi-particle $\delta$-peak appears in the spectral function in addition to the hump present at the energies of the order of plasmon
energy. The particular attention is devoted to the continuous cross-over from the non-Fermi liquid to the Fermi liquid regime by increasing
$t_\perp$. It is shown that the spectral weight of the hump transfers to the quasi-particle as the optical gap in the plasmon dispersion
increases together with $t_\perp$, with the quasi-particle residuum $Z$ behaving like $- (\ln t_{\perp})^{-1}$ in the limit
$t_{\perp}\rightarrow 0$. Our approach is appropriate for the wide range of energy scales given by the plasmon energy
and the width of the conduction band, and is complementary to the Luttinger liquid techniques that are limited to the low energy regime close to
the Fermi surface.
\end{abstract}

%\pacs{}
%\keywords{}

\maketitle

\section{INTRODUCTION}

Recent ARPES measurements of photoemission spectra show that a series of quasi-one-dimensional conductors, in particular the acceptor-donor
chain compound TTF-TCNQ \cite{Zwick1,Claessen} and Bechgaard salts (TMTSF)$_2$X with X = PF$_6$, ClO$_4$, ReO$_4$,...
\cite{Zwick,Zwick2,Zwick3}, have unusual properties, clearly distinguishable from the spectra of standard three-dimensional conductors.
Quasi-particle peaks in these compounds are absent, and the spectra are instead dominated by a wide feature spread across energy scales of the
order of plasmon energies. Such data are in qualitative accordance with the conclusions of our recent calculation~\cite{bbz} for the spectral
function of the one-dimensional electron band with the three dimensional long range Coulomb electron-electron interaction, obtained within the
so-called $G_{0}W_{0}$ approximation \cite{Hedin1}. The physical origin of such behavior is the one-dimensionality of the electron band that
causes an anisotropic acoustic plasmon dispersion. Since such dispersion spreads through the whole range of energies, from zero up
to the plasmon energy $\Omega_{pl}$, it introduces the wide feature into the spectral function at these energies, leaving thus no space for the
creation of quasi-particle $\delta$-peaks.

The spectral density $N(\omega)$ and other quantities related to the electron spectral properties have been also calculated exactly within the
Luttinger liquid approach, using mostly the bosonization method ~\cite{Meden3,Voit2}. Such analyzes are however limited  to the
narrow range of low energies, $\omega\ll E_F, \Omega_{pl}$, where $E_F$ is the Fermi energy of the order of bandwidth. It was shown that,
together with the absence of quasi-particle peaks, the spectral function shows power law behavior with the anomalous dimension $\alpha$, defined
by $N(\omega) \sim |\omega|^{\alpha}$ ~\cite{Meden3,Voit2} and being interaction dependent. The comparison with measurements at low frequencies
suggests values of anomalous dimension in the range $\alpha>1$. This corresponds to the regime of strong three-dimensional long-range Coulomb
interactions ~\cite{Meden1,Barisic,Schulz,Botric,Meden2}, which additionally suggests that the corresponding plasmon energy scale is not small,
being at least of the order of band-width or larger. The $G_{0}W_{0}$ approximation is the only known approach which, as was already pointed
out, enables the calculation of spectral properties in such wide ranges. However, it does not lead to the correct power law exponent in the
limit $\omega\rightarrow 0$. As such, it is complementary to the Luttinger liquid approach ~\cite{Meden3,Voit2} which is concentrated and
limited to the low energy region.

The combination of two above approaches thus covers the whole energy range relevant for the analysis of the photoemission properties of
quasi-one-dimensional metals. As was already stated, the main emerging conclusion for the electron liquid with a strictly one-dimensional band
dispersion is that, although three-dimensionally coupled through long-range Coulomb interaction, it does not show the essential property of
Fermi liquids, namely the presence of quasi-particle excitations in the one-particle spectral properties. However in order to understand better
the spectral properties of real quasi-one-dimensional conductors one has to take into account deviations from the one-dimensional band
dispersion which come from finite inter-chain electron tunnellings. The corresponding question of both, theoretical and experimental interests
is: how one reestablishes the Fermi liquid character of spectral properties by introducing and gradually increasing the transverse bandwidth
$t_\perp$, approaching thus the regime of standard isotropic three-dimensional conducting band?

In this work we address this question by extending our earlier $G_{0}W_{0}$ approach to the rectangular lattice of parallel chains with a finite
transverse tunnelling integral $t_\bot$. After taking into account the corresponding finite transverse curvature in the three-dimensional band
dispersion ~\cite{Kwak,zup}, the screened Coulomb interaction $W_{0}$ calculated within the random phase approximation (RPA) shows a finite
optical plasmon gap proportional to $t_\bot$ in the long-wavelength limit. The plasmon dispersion thus has a three-dimensional, albeit strongly
anisotropic, character for any finite value of $t_\bot$. A more detailed insight into the electron self-energy within the $G_{0}W_{0}$ approach
shows that this property of plasmon dispersion has  the dominant effect on the dressed electron propagator through the screened Coulomb
interaction $W_{0}$, while the influence of finite $t_\bot$ through a bare electron propagator $G_{0}$ can be neglected. This enables an
analytical derivation of the dressed electron Green's function and other quantities that follow from it.

The obtained result reveals the appearance of low energy quasi-particle peaks, in addition to the smeared structure at higher energies which is
a characteristic of the strictly one-dimensional ($t_\bot=0$) limit \cite{bbz}. Note that the early $G_{0}W_{0}$ approach to the isotropic
three-dimensional ,,jellium'' ~\cite{Hedin,Lundq2,Lundq1} led to the analogous result for the spectral function, showing quasi-particle peaks in
the energy range $\mu-\Omega_{pl}<\omega<\mu+\Omega_{pl}$ where $\Omega_{pl}$ is the minimum of the optical long-wavelength plasmon dispersion,
and an additional structure due to the plasmon mode, with the finite spectral weight below and above these energies.

The spectral properties for the generalized Luttinger liquid with a weak electron tunnelling between metallic chains and with the
three-dimensional electron-electron Coulomb interaction were analyzed by using the appropriately developed higher-dimensional bosonization
technique ~\cite{Meden1,Meden2} in which the Fermi surface is approximated by a finite number of flat patches. This technique inherits in itself
two approximations, namely the momentum transfer between different patches is ignored and the local band dispersion is linearized. On the other
hand, it handles the case of $t_\perp\neq 0$ without having to rely on an expansion in powers of $t_\perp$ used in earlier studies of the model
of parallel chains with a finite inter-chain hopping ~\cite{Wen, Bourbonnais5,Boies,Clarke, Tsvelik}. Using the 4-patch approximation for the
Fermi surface Kopietz et al. ~\cite{Meden1,Meden2} obtained in the strong coupling limit the spectral function with the low energy
quasi-particle having the weight proportional to $\Theta^{\gamma_{cb}}$, $\Theta=|t_\perp|/E_{F}$. Here $\gamma_{cb}$ is the
anomalous dimension of corresponding Luttinger liquid for $t_\perp=0$, and $E_{F}$ is Fermi energy. Furthermore, it is shown that there exists a
large intermediate regime of wave vectors and frequencies where the Green's function satisfies the same anomalous scaling behavior as for
$t_\perp=0$. This is to be contrasted with the result of the perturbation treatment of  $t_\perp$ ~\cite{Wen} in which the quasi-particle peak
appears only when the one-dimensional Green's function diverges, i. e. for the anomalous dimension less than unity.

Again, like in the case $t_\perp=0$, the higher dimensional bosonization and our $G_{0}W_{0}$ approach are complementary, since the former is
limited to the scaling behavior of the Green's function in the low energy range and the latter enables the reliable calculation of the wide
maximum at the range of plasmon energy in the spectral function. It is important to note that the essential ingredient in both approaches is
that the finite $t_\perp$ enters into calculations through the long wavelength optical gap in the plasmon dispersion, and not through the
corrugation of the band dispersion at the Fermi energy as in the perturbation approach of Ref.~\cite{Wen}. On the other side, while both Wen's
expansion in terms of  $t_\perp$ \cite{Wen} and the higher-dimensional bosonization treatment cover low energy scaling, only the present
$G_{0}W_{0}$ approach describes appropriately the cross over from the one-dimensional non-Fermi liquid regime to the three-dimensional Fermi
liquid one in the whole range of energies.

In Section II we calculate the electron Green's function within the $G_{0}W_{0}$ method developed in our previous work ~\cite{bbz}. Section III
is devoted to the spectral function. The density and the momentum distribution function are discussed in Section IV. Section V contains
concluding remarks.

\section{GREEN'S FUNCTION}
\subsection{Dielectric function and excitations}
We begin by considering the effect of finite transverse bandwidth on the plasmon dispersion. The electron band dispersion is modeled by \be
\label{disp} E({\bf k})=-2t_{0}(\cos{k_{\parallel}b}-\cos{k_{F}b})-2t_\perp(\cos{k_{x}a}+\cos{k_{z} c}),
 \ee
where $b$ and  $a,c$ are longitudinal and two transverse lattice constants respectively, while $t_{0}$ and $t_\perp$ are corresponding transfer
integrals. The RPA polarization diagram now reads
 \be
 \label{pola16}
 \Pi({\bf q},\omega)= \frac{4}{N_{a}N_{b}N_{c}} \sum_{k_{x}=-\frac{\pi}{a}}^{\frac{\pi}{a}}
\sum_{k_{\parallel}=-\frac{\pi}{b}}^{\frac{\pi}{b}}
 \sum_{k_{z}=-\frac{\pi}{c}}^{\frac{\pi}{c}}\frac{n({\bf k})
[E({\bf k}+{\bf q})-E({\bf k})]}{ (\omega +i\eta \textrm{sign}\omega)^{2} -
  [E({\bf k}+{\bf q})-E({\bf k})]^{2}},
\ee
where
\be
\label{occ} n({\bf k})=\left \{ \begin{array}{ll}
1, & E({\bf k})<E_{F}\\
0, & E({\bf k})>E_{F}
\end{array} \right.
\ee is the occupation function. In the long wave-length limit ${\bf q}\to 0$, where $\omega\gg E({\bf k}+{\bf q})-E({\bf k})$, the polarization
diagram reduces to \cite{Ziman}
\be
\label{pola18}
\Pi({\bf q},\omega)= \frac{2}{N_{a}N_{b}N_{c}(\omega+i\eta \textrm{sign}\omega)^{2}}
\sum_{k_{x}=-\frac{\pi}{a}}^{\frac{\pi}{a}} \sum_{k_{\parallel}=-\frac{\pi}{b}}^{\frac{\pi}{b}}
 \sum_{k_{z}=-\frac{\pi}{c}}^{\frac{\pi}{c}}n({\bf k})
({\bf q}\cdot \nabla_{\bf k})^{2}E({\bf k}) \ee with \be \label{nabla} ({\bf q}\cdot \nabla_{\bf k})^{2}E({\bf
k})=q_{x}^{2}\frac{\partial^{2}E({\bf k})}{\partial k_{x}^{2}}+q_{\parallel}^{2} \frac{\partial^{2}E({\bf k})}{\partial
k_{\parallel}^{2}}+q_{z}^{2}\frac{\partial^{2}E({\bf k})}{\partial k_{z}^{2}}. \ee Since by assumption $t_\perp \ll t_0$, the Fermi surface is
only slightly corrugated, i. e. $\delta(k_x,k_z) / k_{F}\ll 1$, where $\delta(k_x,k_z)$ is the deviation of the component of the Fermi wave
vector in the chain direction from $k_{F}$, the latter being its value at $t_\perp=0$. The expansion of the band dispersion
(\ref{disp}) in terms of $\delta$ up to the second order ~\cite{Kwak} leads to the equation for the Fermi surface \be \label{razvoj}
E(k_{x},k_{F}+\delta,k_{z}) \equiv v_{F}\delta+E_{F}''\delta^{2}/2-2t_\perp(\cos k_{x}a+\cos k_{z} c)= E_F \ee where $v_{F}=2t_{0}b\sin{k_{F}b}$
is Fermi velocity, $E_{F}''\equiv
\partial^{2}E({\bf k})/\partial k_{\parallel}^{2}$ at $k_\parallel =k_{F}$, and $E_F$ is the shift of the Fermi energy with respect to its value
for $t_\perp=0$. Our aim is to find out how $\delta(k_x,k_z)$ depends on $t_\perp$, and to determine the corresponding value of $E_F$. To this
end we note that by switching to finite $t_\perp$ the band filling does not change, so that \be \label{cond}
\int_{-\pi/a}^{\pi/a}\int_{-\pi/c}^{\pi/c}\delta(k_x,k_z)dk_{x}dk_{z}=0 .\ee Then, since $\delta\sim t_\perp$ to the lowest order, the
integration of Eq. \ref{razvoj} in terms of $k_x$ and $k_z$ gives $E_F \sim t_\perp^{2}$. The explicit expansions follow after expressing
$\delta(k_x,k_z)$ from Eq. \ref{razvoj},
\begin{eqnarray}
\delta(k_x,k_z)=-\frac{v_{F}}{E_{F}''}\Bigg\{1\pm\sqrt{1-2\frac{E_{F}''}{v_{F}^{2}}\bigg[-E_{F}-2t_\perp(\cos k_{x}a+\cos
k_{z}c)\bigg]}\Bigg\}\nonumber\\\approx \frac{2t_\perp}{v_{F}}(\cos{k_{x}a}+\cos{k_{z}c)}+\frac{E_{F}}
{v_{F}}-2\frac{E_{F}''}{v_{F}^{3}}t_{\perp}^{2}(\cos{k_{x}a}+\cos{k_{z}c})^{2}.
\end{eqnarray}
Inserting this expression into the condition (\ref{cond}) one gets $E_F=2E_{F}''t_{\perp}^{2}/v_{F}^{2}$, and finally \be
\delta(k_x,k_z)=\frac{2t_\perp}{v_{F}}(\cos{k_{x}a}+\cos{k_{z}c)}+2\frac{E_{F}''}
{v_{F}^{3}}t_{\perp}^{2}-2\frac{E_{F}''}{v_{F}^{3}}t_{\perp}^{2}(\cos{k_{x}a}+\cos{k_{z}c})^{2}. \label{delta} \ee

\begin{figure}
\vspace*{9.5cm} \centering \includegraphics{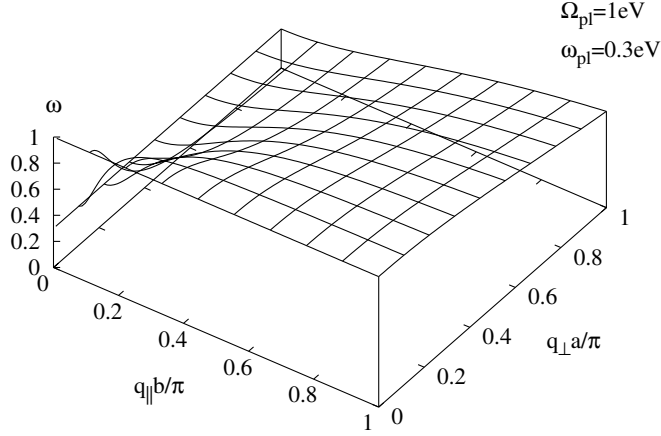} \caption{\label{spektar0}Plasmon dispersion $\omega({\bf
q})$ (see Eq. \ref{plazdisp2}).}
\end{figure}

The expansion (\ref{delta}) enables the analytical derivation of the dielectric function $\varepsilon_{m}({\bf q},\omega)=1 - V({\bf q})\Pi({\bf
q},\omega)$, where $V({\bf q})=\frac{4\pi e^{2}}{v_{0}q^{2}}$ is the bare Coulomb interaction. After replacing
\be
\sum_{k_{x}=-\frac{\pi}{a}}^{\frac{\pi}{a}} \sum_{k_{\parallel}=-\frac{\pi}{b}}^{\frac{\pi}{b}}
 \sum_{k_{z}=-\frac{\pi}{c}}^{\frac{\pi}{c}}n({\bf k})...
\to \bigg(\frac{L}{2\pi}\bigg)^{3} \int_{-\frac{\pi}{a}}^{\frac{\pi}{a}}dk_{x} \int_{-\frac{\pi}{c}}^{\frac{\pi}{c}}dk_{z}
\int_{-(k_{F}+\delta)}^{k_{F}+\delta}dk_{\parallel}...
 \ee
in Eq. \ref{pola18}, and taking into account that
 \be
\int_{-(k_{F}+\delta)}^{k_{F}+\delta}dk_{\parallel}\frac{\partial^{2}E({\bf k})}{\partial k_{\parallel}^{2}}=2\frac{\partial E({\bf k})}
{\partial k_{\parallel}}\bigg\vert_{0}^{k_{F}+\delta}=2(v_{F}+E_{F}''\delta+ E_{F}'''\delta^{2}/2) \ee and \be \frac{\partial^{2}E({\bf
k})}{\partial k_{x}^{2}}=2t_\perp a^{2}\cos k_{x}a, \mbox{ } \frac{\partial^{2}E({\bf k})}{\partial k_{z}^{2}}=2t_\perp c^{2}\cos k_{z}c, \ee we
get \be \varepsilon_{m}({\bf q},\omega)=1-\frac{\omega^{2}({\bf q})}{(\omega +i\eta \textrm{sign}\omega)^{2}}\ee with the plasmon dispersion
given by \be \label{plazdisp1} \omega^{2}({\bf q})=\frac{\Omega_{pl}^{2}q_{\parallel}^{2}+
\omega_{pa}^{2}q_{x}^{2}+\omega_{pc}^{2}q_{z}^{2}}{q^2}. \ee Here longitudinal and transverse plasmon frequencies are given by
$\Omega_{pl}^{2}=\frac{8e^{2}v_{F}}{ac}\Big(1+2\frac{E_{F}'''}{v_{F}^{2}}t_{\perp}^{2}\Big)$ and $\omega_{pa}^{2}=
\frac{16e^{2}t_{\perp}^{2}a}{cv_{F}}$, $\omega_{pc}^{2}= \frac{16e^{2}t_{\perp}^{2}c}{av_{F}}$ respectively. Thus, the finiteness of the
transverse bandwidth retains the optical character of the plasmon dispersion in all directions of the long wavelength range
$\textbf{q}\rightarrow 0$, with the anisotropy scaled by the ratio $\frac{{t_\perp}}{t_0}$ as shown in Fig.~\ref{spektar0}. As long as
$t_{\perp} \ll t_0$ we can skip the correction proportional to $t_{\perp}^{2}$ in $\Omega_{pl}^{2}$. Also, for simplicity we put $a=c$ and get
the simplified expression for the long wavelength plasmon dispersion, \be \label{plazdisp2} \omega^{2}({\bf
q})=\frac{\Omega_{pl}^{2}q_{\parallel}^{2}+ \omega_{pl}^{2}q_{\bot}^{2}}{q^2}, \ee with $\omega_{pl}^{2}= \frac{16e^{2}t_\perp^{2}}{v_{F}}$ and
$q_{\bot}^{2}\equiv q_{x}^{2}+q_{z}^{2}$. Note that in the regime of strong Coulomb interaction, $\Omega_{pl} \gg t_0$ \cite{bbz}, we also have
\be \label{ineq} \frac{\omega_{pl}}{t_\perp} = \sqrt{\frac{ac}{2b^{2}}}\frac{\Omega_{pl}}{t_0 \sin(k_F b)}\gg 1. \ee

\subsection{Green's function}

In the calculation of the reciprocal Green's function $G^{-1}({\bf k},\omega)$, we follow the $G_{0}W_{0}$ approximation \cite{bbz}. The
extension of this procedure obtained by the inclusion of the full ${\bf q}$-dependence in the band dispersion (\ref{disp}) leads to the
generalization of the equation (20) in Ref. \cite{bbz},
\begin{eqnarray}
\label{GP271}
G^{-1}({\bf
k},\omega)=\omega-E({\bf k})+i\eta[1-2n({\bf k})]
-E_{ex}({\bf k})
-\frac{1}{2N}\sum_{\bf q}V({\bf
q})\omega({\bf q})\times {}
\nonumber\\
{}\times\Bigg[\frac{1-n({\bf k}+{\bf q})}{\omega-\mu- \omega({\bf q})-E({\bf k}+{\bf q}) +i\eta} +\frac{n({\bf k}+{\bf q})}
{\omega-\mu+\omega({\bf q})-E ({\bf k}+{\bf q})-i\eta}\Bigg]. {}
\end{eqnarray}
Here \be \label{exch11} E_{ex}({\bf k})=-\frac{1}{N}\sum_{\bf q}V({\bf q})n({\bf k}+{\bf q}) \ee is the exchange energy per elementary cell for
the one particle state with the wave vector ${\bf k}$. Further simplification follows after noticing that, as far as we are in the regime of
strong Coulomb interaction, $\Omega_{pl} \gg t_0$ (see Ref. \cite{bbz} and Eq. \ref{ineq}), two second terms in the dispersion $E({\bf k}+{\bf
q})\approx E_{0}(k_{\parallel})+v_{F}q_{\parallel}+E_{\bot}(k_{a}+q_{a},k_{c}+q_{c})$ appearing in the denominators of Eq. \ref{GP271} can be
neglected with respect to that of the plasmon dispersion $\omega({\bf q})$. As it will be seen later, this approximation introduces
small losses in the spectral density at low frequencies, but does not affect its main qualitative features. After a few nonessential
simplifications which do not affect the physical content, like taking the flat Fermi surface at $\vert k_{\parallel}\vert =k_{F}$ for the
occupation function (\ref{occ}) and using cylindrical coordinates in the integration across the I Brillouin zone \cite{bbz}, one gets the
analytical expression for $G^{-1}({\bf k},\omega)$. Its real part reads
\begin{eqnarray}
\label{re113}
&&\!\!\!\!\!\!\!\!\!\!\!\!ReG^{-1}({\bf k},\omega)=\omega-E({\bf k})
+\frac{e^{2}}{2b}\bigg\{\ln\bigg[\bigg(\frac{bQ_{\bot}}{\pi}\bigg)^{2}+1\bigg]
+\frac{2bQ_{\bot}}{\pi}\arctan\frac{\pi}{bQ_{\bot}}\bigg\}
\nonumber\\
&-&\!\!\!
\frac{e^{2}}{2\pi}\Bigg\{
\frac{(\omega-\mu-E_{0}(k_{\parallel}))\omega_{pl}}{(\omega-\mu-E_{0}(k_{\parallel}))^{2}-\omega_{pl}^{2}}\frac{2\pi}{b}
\Bigg[\ln\frac{\omega_{pl}}{\omega_{pl}+\Omega_{pl}}
+\ln\Bigg\vert\sqrt{1+\Big(\frac{bQ_{\bot}}{\pi}\Big)^{2}}
+\sqrt{\frac{\Omega_{pl}^{2}}{\omega_{pl}^{2}}+\Big(
\frac{bQ_{\bot}}{\pi}\Big)^{2}}\Bigg\vert\Bigg]
\nonumber\\
&+&\!\!\!
\frac{(\omega-\mu-E_{0}(k_{\parallel}))^{2}}{(\omega-\mu-E_{0}(k_{\parallel}))^{2}-\omega_{pl}^{2}}
\Bigg[F\bigg(\frac{\pi}{b},\omega-\mu\bigg)
-R(k_{\parallel},\omega-\mu)
\nonumber\\
&+&\!\!\!\frac{2\pi}{b}\ln\Bigg\vert\frac{(\omega-\mu-E_{0}(k_{\parallel}))
\sqrt{1+\Big(\frac{bQ_{\bot}}{\pi}\Big)^{2}}-\omega_{pl}
\sqrt{\frac{\Omega_{pl}^{2}}{\omega_{pl}^{2}}+\Big(
\frac{bQ_{\bot}}{\pi}\Big)^{2}}}{\omega-\mu-E_{0}(k_{\parallel})
-\Omega_{pl}}\Bigg\vert
\Bigg]
\nonumber\\
&+&\!\!\!Q_{\bot}\frac{(\omega-\mu-E_{0}(k_{\parallel}))\Omega_{pl}^{2}}{\omega_{pl}
((\omega-\mu-E_{0}(k_{\parallel}))^{2}-\Omega_{pl}^{2})}
\int_{-\frac{\pi}{bQ_{\bot}}}^{\frac{\pi}{bQ_{\bot}}}
\frac{dy}{\sqrt{y^{2}+1}\sqrt{\frac{\Omega_{pl}^{2}}{\omega_{pl}^{2}}y^{2}+1}}
\\
&+&\!\!\!Q_{\bot}\frac{(\omega-\mu-E_{0}(k_{\parallel}))^{3}(\omega_{pl}^{2}-\Omega_{pl}^{2})}
{\omega_{pl}((\omega-\mu-E_{0}(k_{\parallel}))^{2}-\Omega_{pl}^{2})^{2}}
\int_{-\frac{\pi}{bQ_{\bot}}}^{\frac{\pi}{bQ_{\bot}}}
\frac{dy}{\sqrt{y^{2}+1}\sqrt{\frac{\Omega_{pl}^{2}}{\omega_{pl}^{2}}y^{2}+1}
\Big[y^{2}+\frac{(\omega-\mu-E_{0}(k_{\parallel}))^{2}-\omega_{pl}^{2}}
{(\omega-\mu-E_{0}(k_{\parallel}))^{2}-\Omega_{pl}^{2}}\Big]}
\Bigg\}
\nonumber{}
\end{eqnarray}
with functions $R$ and $F$ given by the expressions
\begin{eqnarray}
\label{R}
R(k_{\parallel},\omega)&=&
\bigg[R_{1}(k_{F}-|k_{||}|,\omega)+R_{1}(k_{F}+|k_{\parallel}|,\omega)\bigg]
\Theta\bigg(\frac{\pi}{b}-|k_{\parallel}|-k_{F}\bigg) {}\\
&+&\bigg[R_{1}(k_{F}-|k_{||}|,\omega)+2R_{1}\bigg(\frac{\pi}{b},\omega\bigg)
-R_{1}\bigg(\frac{2\pi}{b}-k_{F}-|k_{\parallel}|,\omega\bigg)\bigg]
\Theta\bigg(k_{F}+|k_{\parallel}|-\frac{\pi}{b}\bigg), {}
\nonumber
\end{eqnarray}
with \be
R_{1}(x,\omega)=\left \{ \begin{array}{ll} -2x\ln\mid x\mid+x\ln\Bigg\vert x^{2}+Q_{\bot}^{2}\frac{(\omega-E_{0}(k_{\parallel}))^{2}-\omega_{pl}^{2}}
{(\omega-E_{0}(k_{\parallel}))^{2}-\Omega_{pl}^{2}}\Bigg\vert +F(x,\omega), & x\neq 0, \\
0, x=0 \end{array} \right. \ee  and \be \label{F} \!\!F(x,\omega)\!\!=\!\!\left \{ \begin{array}{ll}2
Q_{\bot}\sqrt{\frac{(\omega-E_{0}(k_{\parallel}))^{2}-\omega_{pl}^{2}} {(\omega-E_{0}(k_{\parallel}))^{2}-\Omega_{pl}^{2}}}\times
\arctan\frac{x} {Q_{\bot}\sqrt{\frac{(\omega-E_{0}(k_{\parallel}))^{2}-\omega_{pl}^{2}} {(\omega-E_{0}(k_{\parallel}))^{2}-\Omega_{pl}^{2}}}}
\,\,\,  for  \,\,\,
|\omega-E_{0}(k_{\parallel})|\!\!<\!\!\omega_{pl},\Omega_{pl}\!\!<\!\!|\omega-E_{0}(k_{\parallel})|, \\
Q_{\bot}\sqrt{\frac{ (\omega-E_{0}(k_{\parallel}))^{2}-\omega_{pl}^{2}}
{\Omega_{pl}^{2}-(\omega-E_{0}(k_{\parallel}))^{2}}}\times \ln\Bigg\vert\frac{x+Q_{\bot}\sqrt{\frac{(\omega-E_{0}(k_{\parallel}))^{2}-
\omega_{pl}^{2})}
 {\Omega_{pl}^{2}-(\omega-E_{0}(k_{\parallel}))^{2}}}}
 {x-Q_{\bot}\sqrt{\frac{(\omega-E_{0}(k_{\parallel}))^{2}-\omega_{pl}^{2}}
{\Omega_{pl}^{2}-
 (\omega-E_{0}(k_{\parallel}))^{2}}}}\Bigg\vert \,\,\,  for  \,\,\,
\omega_{pl}\!\!<\!\!|\omega-E_{0}(k_{\parallel})|\!\!<\!\!\Omega_{pl}.
\end{array} \right.
\ee
The exchange energy in the expression (\ref{re113}) is given by
\begin{eqnarray}
 \label{exchapp}
 E_{ex}(k_{\parallel})=- \frac{e^{2}}{2\pi}\Bigg\{\bigg[
 H(k_{F}-|k_{||}|)+H(k_{F}+|k_{\parallel}|)\bigg]
 \Theta\bigg(\frac{\pi}{b}-|k_{\parallel}|-k_{F}\bigg)+
 \nonumber\\
 +\bigg[H(k_{F}-|k_{||}|)+2H\bigg(\frac{\pi}{b}\bigg)
 -H\bigg(\frac{2\pi}{b}-k_{F}-|k_{\parallel}|\bigg)\bigg]
 \Theta\bigg(k_{F}+|k_{\parallel}|-\frac{\pi}{b}\bigg)\Bigg\}
 \end{eqnarray}
 with
 \be
 H(x)\equiv x\ln(Q_{\bot}^{2}+x^{2})+2Q_{\bot}\arctan\frac{x}{Q_{\bot}}-x\ln x^{2}.
 \ee
The further simplification follows after realizing that in the regime of strong Coulomb interaction the self energy contribution is dominant in
comparison to the transverse dispersion term $2t_\perp(\cos{k_{x}a}+\cos{k_{z} c})$ in $E({\bf k})$. Consequently, we can skip the dependence of
$ReG^{-1}({\bf k},\omega)$ on $k_x$ and $k_z$ as irrelevant for further considerations. Namely, after taking into account that
$Q_{\bot}=2\sqrt{\pi}/\sqrt{ac}\ll\pi/b$, the leading contribution to the third term on the right hand side in Eq. \ref{re113} reduces to
$\frac{e^{2}}{2}Q_{\bot} \approx 0.16\frac{\omega_{pl}\Omega_{pl}}{t_{\bot}}\ll t_{\bot}$. This justifies the above approximation, after which
we can proceed to a great extend along the lines of Ref. \cite{bbz}. In particular, the chemical potential $\mu$ in Eq. \ref{re113} is now,
after taking into account the self-consistent condition $Re G^{-1}(k_F, \mu) = 0$, given by \be
\mu=-\frac{e^{2}}{2b}\bigg\{\ln\bigg[\bigg(\frac{bQ_{\bot}}{\pi}\bigg)^{2}+1\bigg] +\frac{2bQ_{\bot}}{\pi}\arctan\frac{\pi}{bQ_{\bot}}\bigg\}.
\ee

The imaginary part of the reciprocal Green's function is given by
\begin{eqnarray}
\label{im113}
&&\!\!\!\!\!\!\!\!\!\!\!Im G^{-1}(k_{\parallel},\omega)=
\frac{e^{2}}{2}\frac{(\omega-\mu-E_{0}(k_{\parallel}))^2}{(\omega-\mu-E_{0}(k_{\parallel}))^2-\omega_{pl}^{2}}\bigg\{2q_{c}
\Theta(\omega-\mu-E_{0}(k_{\parallel}))\!\!-\!\!\Big[\Theta(-\omega+\mu+E_{0}(k_{\parallel}))
\nonumber\\
&&\!\!\!\!\!\!
+\Theta(\omega-\mu-E_{0}(k_{\parallel}))\Big]\times
\Big[2q_{c}\Theta(k_{F}-\vert k_{\parallel}\vert-q_{c})
+2k_{F}\Theta(q_{c}-\vert k_{\parallel}\vert-k_{F})
\nonumber\\
&&\!\!\!\!\!\!+(k_{F}-\vert k_{\parallel}\vert+q_{c})\Theta(\vert
k_{\parallel}\vert+q_{c}-k_{F})
\Theta(k_{F}-\mid \vert
k_{\parallel}\vert-q_{c}\mid)\Theta(\frac{2\pi}{b}-k_{F}
-\vert k_{\parallel}\vert-q_{c})
\nonumber\\
&&\!\!\!\!\!\!+(2k_{F}+2q_{c}-\frac{2\pi}{b})
\Theta(k_{F}-\mid \vert k_{\parallel}\vert-q_{c}\mid)
\Theta(-\frac{2\pi}{b}+k_{F}+
\vert k_{\parallel}\vert+q_{c})
\Big]\bigg\}    {}
\end{eqnarray}
for $\omega_{pl}<|\omega-\mu-E_{0}(k_{\parallel})|<\Omega_{pl}$, and $ImG^{-1}(k_{\parallel},\omega)=0$ elsewhere. The wave number $q_{c}$ in
Eq. \ref{im113} is defined by \be \label{qc} q_{c}=min\bigg(Q_{\bot}\sqrt{\frac{(\omega-\mu-E_{0}(k_{\parallel}))^{2}-\omega_{pl}^{2}}
{\Omega_{pl}^{2}-(\omega-\mu-E_{0}(k_{\parallel}))^{2}}},\frac{\pi}{b}\bigg). \ee
\begin{figure}
\vspace*{7.5cm} \centering \includegraphics{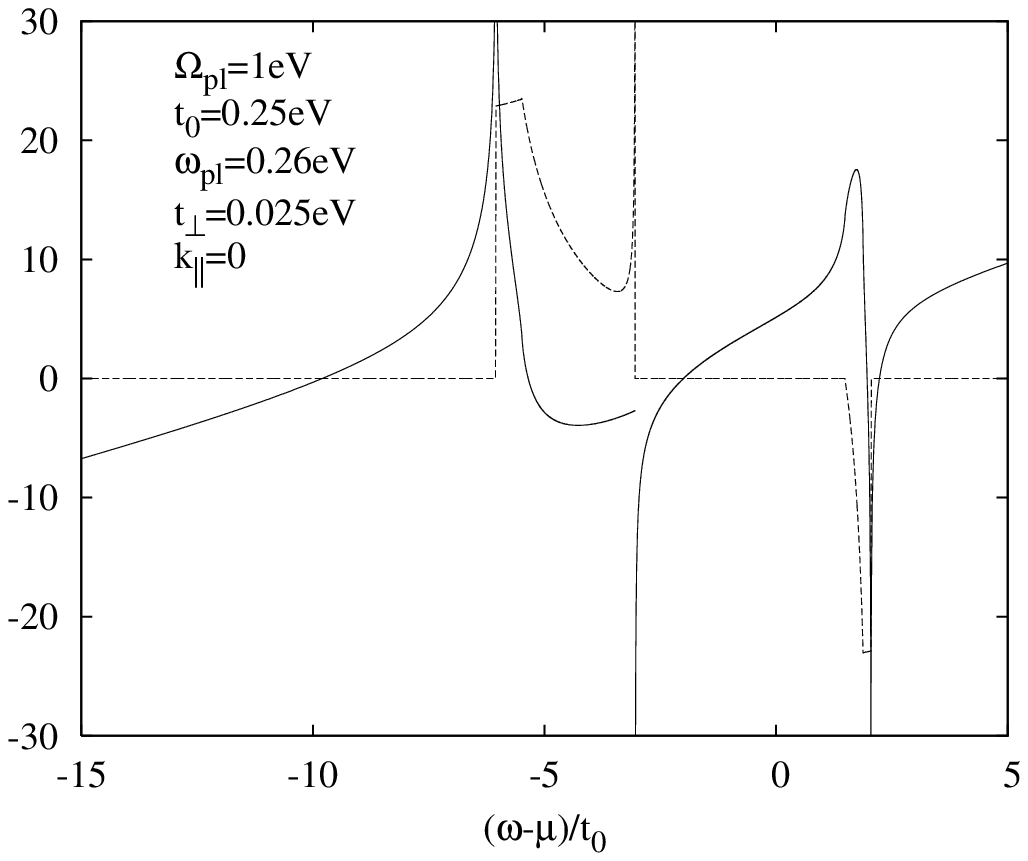} \includegraphics{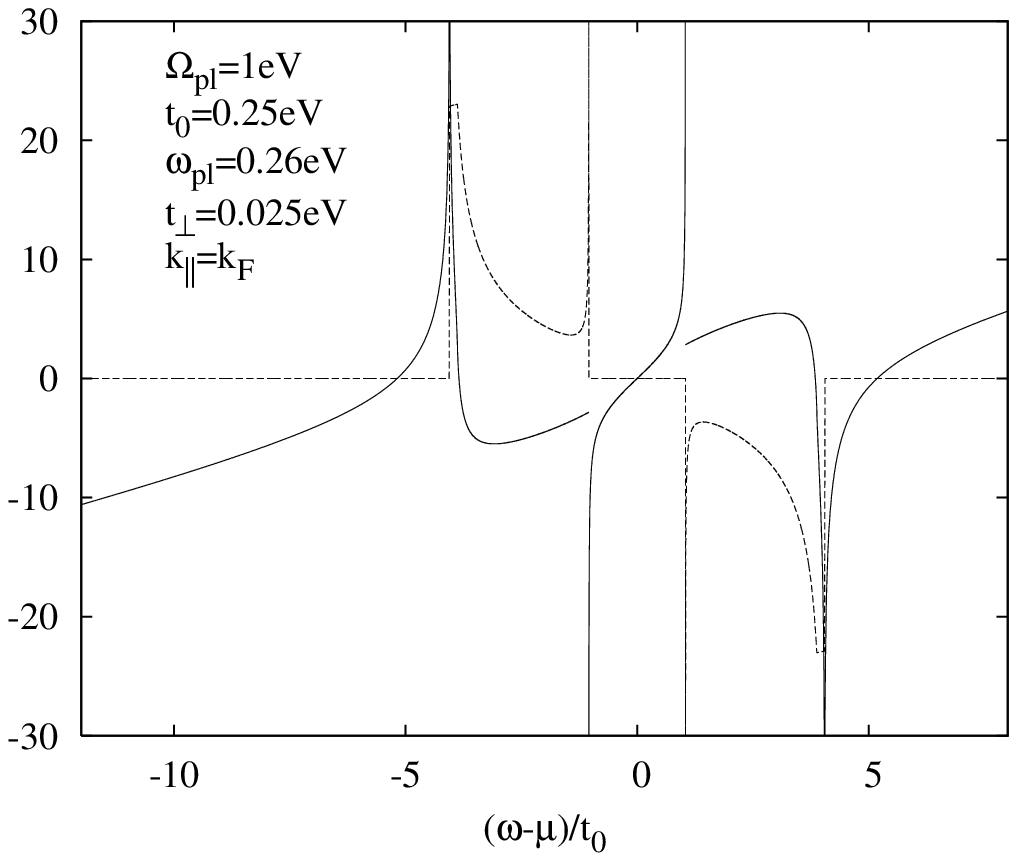}
\includegraphics{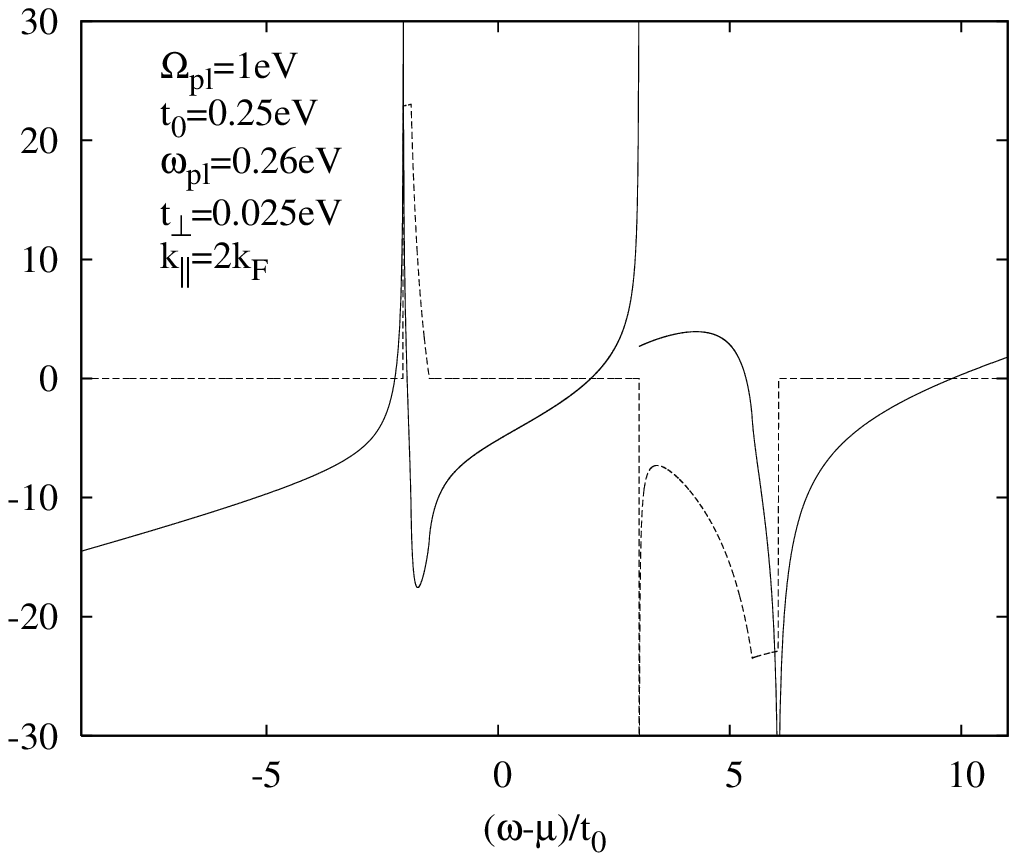}\hbox{\hspace{0.8in} (a) \hspace{2.0in}
(b)\hspace{2.05in} (c)}
\caption{\label{slire1} Frequency dependence of $ReG^{-1}(k_{\parallel},\omega)/t_{0}$ (full lines) and $-ImG^{-1}(k_{\parallel},\omega)/t_{0}$
(dashed lines) for $k_{F}=\pi/2b$ and $k_{\parallel}=0$ (a), $k_{\parallel}=k_{F}$ (b), and $k_{\parallel}=2k_{F}$ (c).}
\end{figure}

$ReG^{-1}(k_{\parallel},\omega)$ and $ImG^{-1}(k_{\parallel},\omega)$ are shown in Fig. \ref{slire1} for three representative values of
$k_{\parallel}$, namely for $k_{\parallel}$ equal to $0$, $k_{F}$, and  $2k_{F}$. Let us at first look more closely into
$ImG^{-1}(k_{\parallel},\omega)$. The vanishing of $ImG^{-1}(k_{\parallel},\omega)$ in the ranges
$|\omega-\mu-E_{0}(k_{\parallel})|<\omega_{pl}$ and $|\omega-\mu-E_{0}(k_{\parallel})|>\Omega_{pl}$ can be traced already from the expression
(\ref{GP271}) after approximating $E({\bf k}+{\bf q})\approx E_{0}(k_{\parallel})$. Namely, in this ranges there are no poles of the reciprocal
Green's function contributing to $ImG^{-1}(k_{\parallel},\omega)$.

$ImG^{-1}(k_{\parallel},\omega)$ vanishes also in the range $\mu+\omega_{pl}+
E_{0}(k_{\parallel})<\omega<\mu+\omega(k_{\parallel}-k_{F},Q_{\bot})+E_{0}(k_{\parallel})$ for $k_{\parallel}< k_{F}$, as well as in the range
$\mu-\omega(k_{\parallel}-k_{F},Q_{\bot})+E_{0}(k_{\parallel})<\omega<\mu-\omega_{pl}+E_{0}(k_{\parallel})$ for $k_{\parallel}> k_{F}$. This
vanishing can be also traced from the expression (\ref{GP271}). Namely, due to the presence of the occupation function $n({\bf k}+{\bf q})$ in
the $\textbf{q}$ - summation the non-vanishing contributions from dense discrete poles at $\omega=\mu-\omega({\bf
q})+E_{0}(k_{\parallel})+i\eta$  contribute only in the range
$\mu-\Omega_{pl}+E_{0}(k_{\parallel})<\omega<\mu-\omega(k_{\parallel}-k_{F},Q_{\bot})+E_{0}(k_{\parallel})$, while the non-vanishing
contributions from poles at $\omega=\mu+\omega({\bf q})+E_{0}(k_{\parallel})-i\eta$ contribute only in the range
$\mu+\omega(k_{\parallel}-k_{F},Q_{\bot})+E_{0}(k_{\parallel})<\omega<\mu+\Omega_{pl}+E_{0}(k_{\parallel})$.

In the range $\omega_{pl}<|\omega-\mu-E_{0}(k_{\parallel})|<\Omega_{pl}$ $ImG^{-1}(k_{\parallel},\omega)$ is covered by the expression
(\ref{im113}). It has a step singularity of the width $\frac{e^2k_{F}\Omega_{pl}^2}{\Omega_{pl}^2-\omega_{pl}^2}$ at
$\omega=\mu\pm\Omega_{pl}+E_{0}(k_{\parallel})$ and diverges at the energies $\omega=\mu-\omega_{pl}+E_{0}(k_{\parallel})$ for
$k_{\parallel}\leq k_{F}$ and $\omega=\mu+\omega_{pl}+E_{0}(k_{\parallel})$ for $k_{\parallel}\geq k_{F}$. At energies
$\omega_{1,2}=\mu\mp\omega(\pi/b,Q_{\bot})+E_{0}(k_{\parallel})$ $Im G^{-1}(k_{\parallel},\omega)$ has respective anomalous minimum and maximum,
with jumps in the first derivatives. These extrema originate from the confinement of the $\textbf{q}$ - summation in the expression
(\ref{GP271}) to the first Brillouin zone. The integration in terms of $q_{\bot}$ from $0$ to $Q_{\bot}$ results in the limitation on the
$q_{\parallel}$ - integration  to the range $|q_{\parallel}|<Q_{\bot}\sqrt{\frac{(\omega-\mu-E_{0}(k_{\parallel}))^{2}-\omega_{pl}^{2}}
{\Omega_{pl}^{2}-(\omega-\mu-E_{0}(k_{\parallel}))^{2}}}$ as far as this limit is within the I Brillouin zone. However, for values of $\omega$
in the ranges $\mu-\Omega_{pl}+E_{0}(k_{\parallel}) <\omega<\mu-\omega(\pi/b,Q_{\bot})+E_{0}(k_{\parallel})$ and
$\mu+\omega(\pi/b,Q_{\bot})+E_{0}(k_{\parallel})<\omega<\mu+ \Omega_{pl}+E_{0}(k_{\parallel})$ we have
$\frac{\pi}{b}<Q_{\bot}\sqrt{\frac{(\omega-\mu-E_{0}(k_{\parallel}))^{2}-\omega_{pl}^{2}}
{\Omega_{pl}^{2}-(\omega-\mu-E_{0}(k_{\parallel}))^{2}}}$, so that the $q_{\parallel}$ - integration is limited to the I Brillouin zone, i. e.
by the $\omega$ - independent boundary $q_{c}=\frac{\pi}{b}$. The resulting values of $Im G^{-1}(k_{\parallel},\omega)$ at the anomalous minimum
and maximum are $\mp e^{2}k_{F}\frac{(\omega_{1,2}-\mu-E_{0}(k_{\parallel}))^{2}}{(\omega_{1,2}-\mu-E_{0}(k_{\parallel}))^{2}
-\omega_{pl}^{2}}$.

Let us now consider $ReG^{-1}(k_{\parallel},\omega)$. As is seen from Fig. \ref{slire1}, it  diverges towards $\pm\infty$ at the
respective energies $\omega=\mu\mp\Omega_{pl}+E_{0}(k_{\parallel})$ at which $ImG^{-1}(k_{\parallel},\omega)$ has step singularities. These
singularities are shifted towards larger values of $\omega$ as $k_{\parallel}$ increases. The zeroes of $ReG^{-1}(k_{\parallel},\omega)$ at
$\omega<\mu-\Omega_{pl}+E_{0}(k_{\parallel})$ and $\omega>\mu+\Omega_{pl}+E_{0}(k_{\parallel})$ are also shifted to the right as
$k_{\parallel}$ increases, former approaching the singularity at $\omega=\mu-\Omega_{pl}+E_{0}(k_{\parallel})$ and latter increasing the
distance from the singularity at $\omega=\mu+\Omega_{pl}+E_{0}(k_{\parallel})$. $ReG^{-1}(k_{\parallel},\omega)$ has also  essential
singularities at $\omega=\mu-\omega_{pl}+E_{0}(k_{\parallel})$ (for $k_{\parallel}\leq k_{F}$) and $\omega=\mu+\omega_{pl}+E_{0}(k_{\parallel})$
(for $k_{\parallel}\geq k_{F}$), i. e. at energies at which $ImG^{-1}(k_{\parallel},\omega)$ diverges.

\begin{figure}
\vspace*{9.5cm} \centering \includegraphics{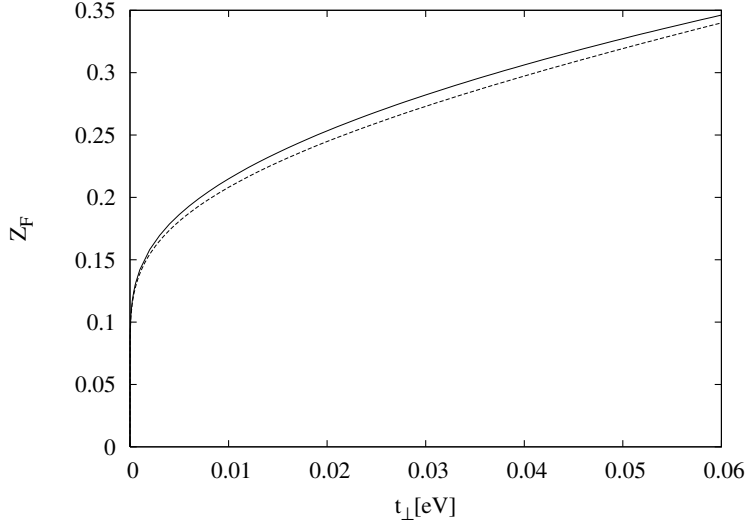} \caption{\label{Zweight} $Z_{F}$ obtained numerically
(full curve) and from the expression (\ref{qweight}) (dashed curve).}
\end{figure}

The zero of $ReG^{-1}(k_{\parallel},\omega)$ in the range $\mu-\omega_{pl}+E_{0}(k_{\parallel})<\omega<\mu+\omega_{pl}+E_{0}(k_{\parallel})$ in
which  $ImG^{-1}(k_{\parallel},\omega)$ vanishes is the low energy pole of the electron propagator $G(k_{\parallel},\omega)$. It is of the form
$y(k_{\parallel})= \widetilde{E}(k_{\parallel})-i\Gamma(k_{\parallel})$, where $\Gamma(k_{\parallel})$ is infinitesimally small in the present
approach. Accordingly, our Green's function has in this range the standard resonant form
 \be
 \label{oblik}
 G(k_{\parallel},\omega)=\frac{Z(k_{\parallel})}{\omega-y
 (k_{\parallel})},
 \ee
where $Z(k_{\parallel}) = \mid \partial ReG^{-1}(k_{\parallel} ,y(k_{\parallel}))/\partial\omega \mid^{-1}$ is the residuum of the Green
function at the pole $y(k_{\parallel})$. We emphasize that the low energy pole appears due to the optical gap $\omega_{pl}$ in the long
wavelength plasmon dispersion introduced by the finite interchain transfer integral $t_{\bot}$ in the electron dispersion. This is illustrated
by the analytical expression for the residuum $Z(k_{\parallel})$ at $k_{\parallel}= k_F$ in the limit $\omega_{pl}\ll\Omega_{pl}$, \be
\label{qweight} Z_{F}=1\bigg/\bigg[1+\frac{e^{2}}{\pi}\frac{Q_{\bot}}{\Omega_{pl}}\ln\bigg(\frac{4\Omega_{pl}}{\omega_{pl}}\bigg)\bigg]=
1\bigg/\bigg[1+\frac{e^{2}}{\pi}\frac{Q_{\bot}}{\Omega_{pl}}\ln\bigg(\frac{\Omega_{pl}\sqrt{v_F}}{et_{\bot}}\bigg)\bigg]. \ee The dependence of
$Z_F$ on $t_{\bot}$ obtained numerically, as well as with the use the expression (\ref{qweight}), is shown in Fig. \ref{Zweight}.
The Green's function has the standard resonant form (\ref{oblik}) also in the frequency range $|\omega-\mu-E_{0}(k_{\parallel})|>\Omega_{pl}$ in
which $ReG^{-1}(k_{\parallel},\omega)$ has zeroes and $ImG^{-1}(k_{\parallel},\omega)$ vanishes.

On the other hand the structure of the Green's function in the region $\omega_{pl}<|\omega-\mu-E_{0}(k_{\parallel})|<\Omega_{pl}$ in which
$ImG^{-1}(k_{\parallel},\omega)\neq 0$ is influenced  by the plasmon dispersion contribution to the expression (\ref{GP271}).

\section{Spectral function}

The single-particle spectral function is defined by \be A(k_{\parallel},\omega)=\frac{1}{\pi}\mid ImG(k_{\parallel},\omega)\mid. \ee It can be
directly expressed in terms of $ReG^{-1}(k_{\parallel}, \omega)$ and $ImG^{-1}(k_{\parallel},\omega)$, \be \label{spp}
A(k_{\parallel},\omega)=\frac{1}{\pi}\frac{\mid ImG^{-1}(k_{\parallel},\omega)\mid}{[ReG^{-1}(k_{\parallel},
\omega)]^{2}+[ImG^{-1}(k_{\parallel},\omega)]^{2}}, \ee unless in the case of $ReG^{-1}(k_{\parallel},\omega)$ having a zero $y(k_{\parallel})$
in the frequency range in which $ImG^{-1}(k_{\parallel},\omega)=0$, when it is represented by the quasi-particle $\delta$-peak \be
\label{izraz1} A(k_{\parallel},\omega)=Z(k_{\parallel})\delta(\omega-y (k_{\parallel})). \ee
\begin{figure}
\vspace*{9.5cm} \centering \includegraphics{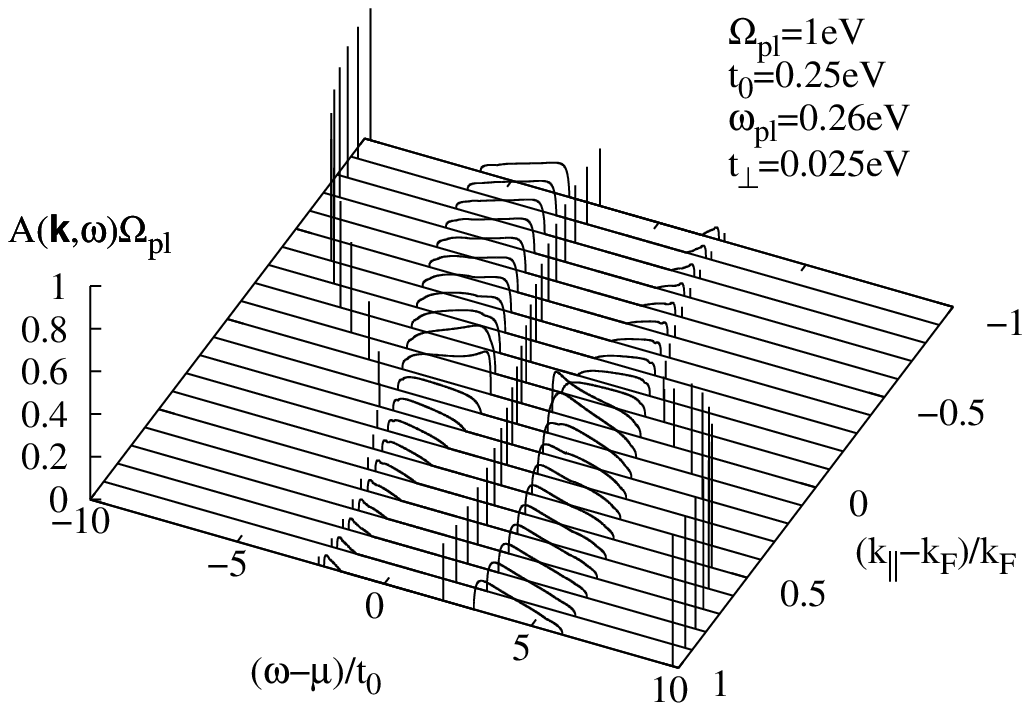} \includegraphics{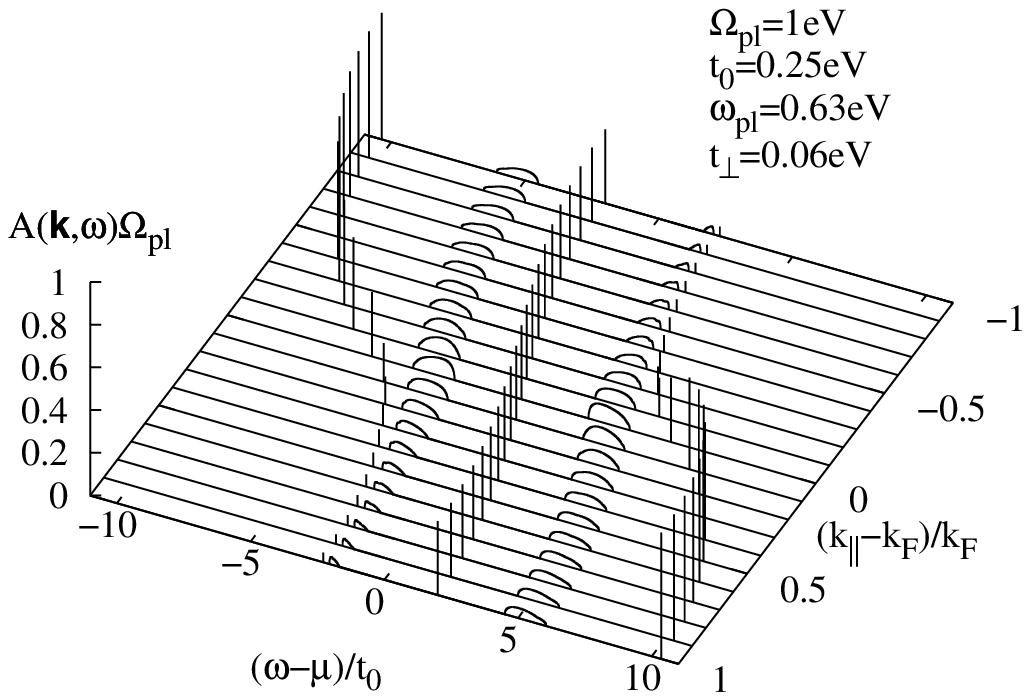}\hbox{\hspace{1.35in} (a) \hspace{2.95in} (b)} \caption{\label{spektar} Spectral function $A(k_{\parallel},\omega)$ for small
($\omega_{pl}=0.26eV$) (a) and large  ($\omega_{pl}=0.63eV$) (b) value of the transverse plasmon frequency $\omega_{pl}$ in the case
$k_{F}=\pi/2b$. Broad maxima for different values of the wave number $k_{\parallel}$ follow from Eq.(\ref{spp}), while $\delta$-peaks are
represented by their weight $Z(k_{\parallel})$ according to Eq. \ref{izraz1}.}
\end{figure}

The spectral function $A(k_{\parallel},\omega)$, obtained after inserting expressions (\ref{re113}) and (\ref{im113}) into Eqs. (\ref{spp}) and
(\ref{izraz1}), is shown in Fig. \ref{spektar} for two values of the transverse plasmon frequency,  $\omega_{pl}=0.26eV$ and $0.63eV$. Generally
it is characterized by the coexistence of wide humps and quasi-particle $\delta$-peaks. Humps originate from the plasmon dispersion in the range
$\omega_{pl}<|\omega-\mu-E_{0}(k_{\parallel})|<\Omega_{pl}$. Their positions vary slowly with the wave number $k_{\parallel}$. As for the
$\delta$-peaks, they are situated in the energy ranges $\mu+E_{0}(k_{\parallel})-\omega_{pl}<\omega<\mu+E_{0}(k_{\parallel})+\omega_{pl}$ and
$|\omega-\mu-E_{0}(k_{\parallel})|>\Omega_{pl}$. It is to be noted that $\delta$-peaks are present for any finite $t_\perp$. However, the
decrease of $t_\perp$ leads to the decrease of the weight of the quasi-particle $\delta$-peak in the range
$\mu+E_{0}(k_{\parallel})-\omega_{pl}<\omega<\mu+E_{0}(k_{\parallel})+\omega_{pl}$ in favor of the growing weight of the hump. In the limit
$t_\perp\to0$, i. e. $\omega_{pl}\to0$, these quasi-particles disappear and all their spectral weight transfers to the hump. The vanishing of
the quasi-particle weight in the range $\mu+E_{0}(k_{\parallel})-\omega_{pl}<\omega<\mu+E_{0}(k_{\parallel})+\omega_{pl}$ as $t_\perp\to0$ is
visible in the dependence of $Z(k_{\parallel})$ on $t_\perp$ for $k_{\parallel}= k_F$ as shown by Eq. \ref{qweight} and in Fig. \ref{Zweight}. We thus come to the spectral function for $t_\perp=0$ which has no low energy quasi-particle. In other words, the
cross-over from the $t_\perp\neq 0$ Fermi liquid regime to the $t_\perp = 0$ non-Fermi liquid regime takes place through the decrease of the
quasi-particle weight by closing the optical gap in the long wavelength plasmon mode.

We note that  numerically obtained spectral function shown in Fig. \ref{spektar} fulfils excellently the sum rule \be
\int_{-\infty}^{\infty}A(k_{\parallel},\omega)d\omega=1, \ee with the agreement up to $10^{-4}$ in the whole range of the wave vector
$k_{\parallel}$, and for all considered values of $t_\perp$. Finally, we notice that, in contrast to the quasi-particles in the
range $\mu+E_{0}(k_{\parallel})-\omega_{pl}<\omega<\mu+E_{0}(k_{\parallel})+\omega_{pl}$, the quasi-particles in the energy range
$|\omega-\mu-E_{0}(k_{\parallel})|>\Omega_{pl}$ are not critically sensitive to the plasmon optical gap $\omega_{pl}$ and keep a finite
intensity in the limit $t_\perp\to0$ as was already shown in Ref. \cite{bbz}.

As was already mentioned in the Introduction, the main property of the above spectral function, namely the quasi-particles at low energies
coexisting with the wide structure originating from the collective plasmon branch,  resembles to the result obtained in the early investigation
of the isotropic ,,jellium'' model within the $G_{0}W_{0}$ approach by Hedin and Lundqvist \cite{Hedin,Lundq2,Lundq1}. They showed that due to
the finite long-wavelength minimum in the optical plasmon dispersion, $\Omega_{pl}$, a quasi-particle with reduced weight appears in the region
$\mu-\Omega_{pl}<\omega<\mu+\Omega_{pl}$, while the rest of the spectral weight is widely distributed at energies outside this range.

As was already argued in Ref. \cite{bbz}, the non-Fermi liquid regime for $t_\perp=0$ is in the qualitative agreement with the ARPES spectra of
Bechgaard salts which apparently do not show low energy quasi-particles ~\cite{Zwick,Zwick2,Zwick3}. On the other hand, the present results for
the spectral function of the quasi-one-dimensional metal in the $t_\perp\neq 0$ Fermi liquid regime suggest that in (TMTSF)$_{2}$PF$_{6}$ (for
which $t_\perp=0.0125eV$ and $t_{0}=0.125eV$) the quasi-particle $\delta$-peak with the weight of the order of $20\%$ of the total spectral
weight for a given value of $k_{\parallel}$ is expected in the low energy range, at an energy distance of the order of
$\omega_{pl}=0.13eV$ from the lower edge of the wide hump. A more directed experimental search, supported by improved energy and intensity
resolutions, is very probably necessary for finding peaks with so weak intensities.

Finally, we refer to the work \cite{Artemenko} devoted to the quasi-two-dimensional metals with the finite transverse transfer integral
$t_\perp$ between metallic planes, with the main result analogous to ours. Namely the spectral function in this case also consists of the
suppressed quasi-particle peak and a broad feature.  Again, the RPA screened Coulomb interaction gives a strongly anisotropic plasmon branch
dispersion of the form (\ref{plazdisp2}) containing small transverse plasmon frequency compared with the longitudinal one. This result is in
agreement with the ARPES spectra of quasi-two-dimensional high-T$_c$ superconductors in the normal conducting phase \cite{Dessau}.

\section{Density of states and momentum distribution function}

\begin{figure}
\vspace*{9.5cm}
\includegraphics{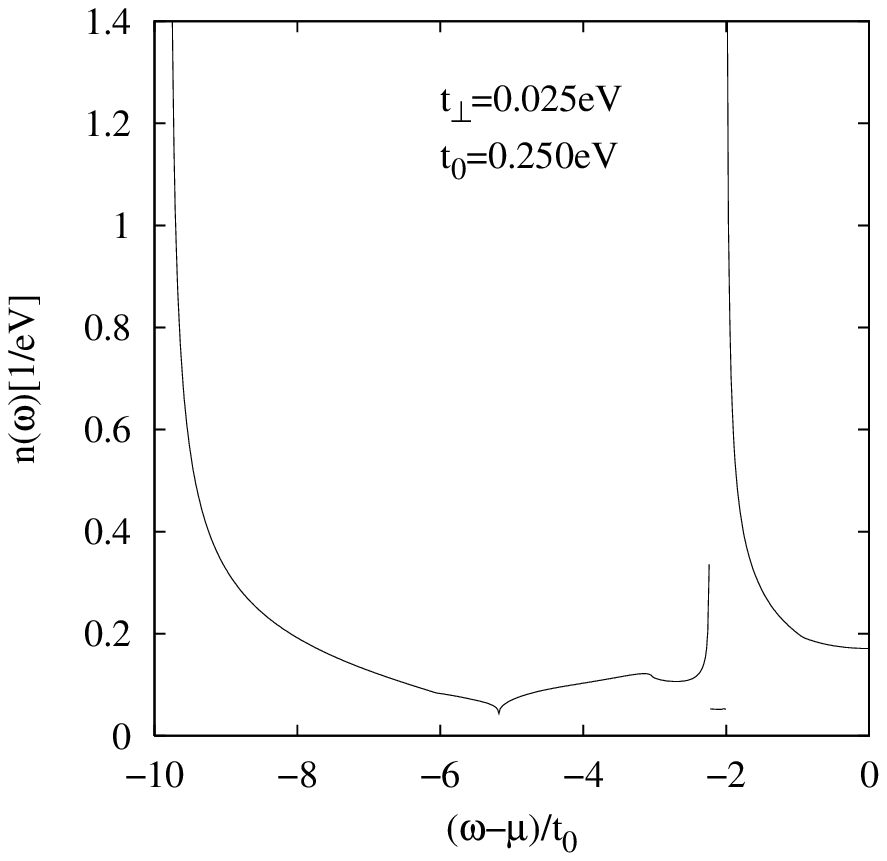}
\includegraphics{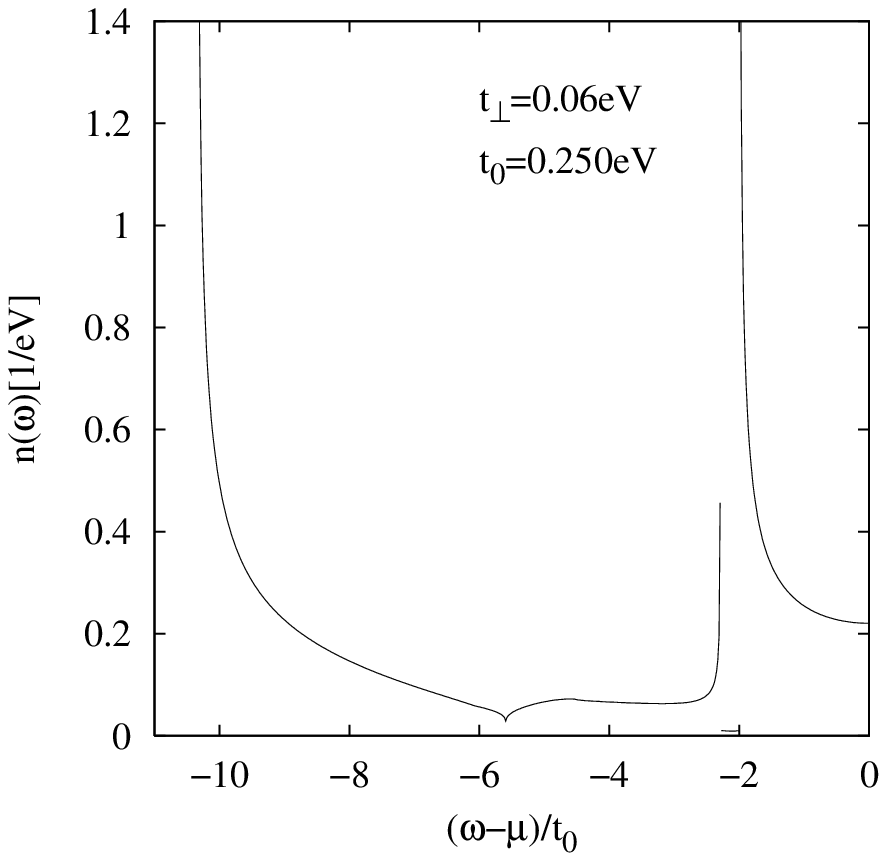} \hbox{\hspace{1.35in} (a) \hspace{3.0in} (b)}
\caption{\label{nodw}Density of states $n(\omega)$ for $t_\perp$ equal $0.025eV$ (a) and $0.06eV$ (b).}
\end{figure}

Integrating numerically the spectral density $A(k_{\parallel},\omega)$ in terms of $k_{\parallel}$, we get the density of states for band
electrons, \be \label{density1} n(\omega)=\frac{1}{2k_{F}}\int_{0}^{\frac{\pi}{b}} A(k_{\parallel},\omega)dk_{\parallel}, \ee shown in Fig.
\ref{nodw} for two values of interchain transfer integral, $t_\perp = 0.025eV$ and $0.06eV$. Three distinctive step singularities in $n(\omega)$
originate from the edges of the corresponding quasi-particle $\delta$-peak dispersions. In particular, the density of states falls from a
maximum at the lowest energy of the $k_{\parallel}$-dependent quasi-particle $\delta$-peak in the range $\omega
<\mu+E_{0}(k_{\parallel})-\Omega_{pl}$ to a local minimum. Then it rises until the step discontinuity at the highest energy of the
quasi-particle $\delta$-peak in the energy range $\omega <\mu+E_{0}(k_{\parallel})-\Omega_{pl}$. Further on, $n(\omega)$ varies slowly from this
discontinuity until the next one at the lowest energy of the quasi-particle $\delta$-peak in the energy range
$\mu+E_{0}(k_{\parallel})-\omega_{pl}<\omega<\mu+E_{0}(k_{\parallel}) +\omega_{pl}$, accumulating the contribution from the spectral density
hump in this range. Increasing further the energy above the third step discontinuity one comes to the minimum of $n(\omega)$ at $\omega=\mu$,
the latter bearing the contribution from the quasi-particle at the chemical potential in the spectral function.

\begin{figure}
\vspace*{9.5cm}
\includegraphics{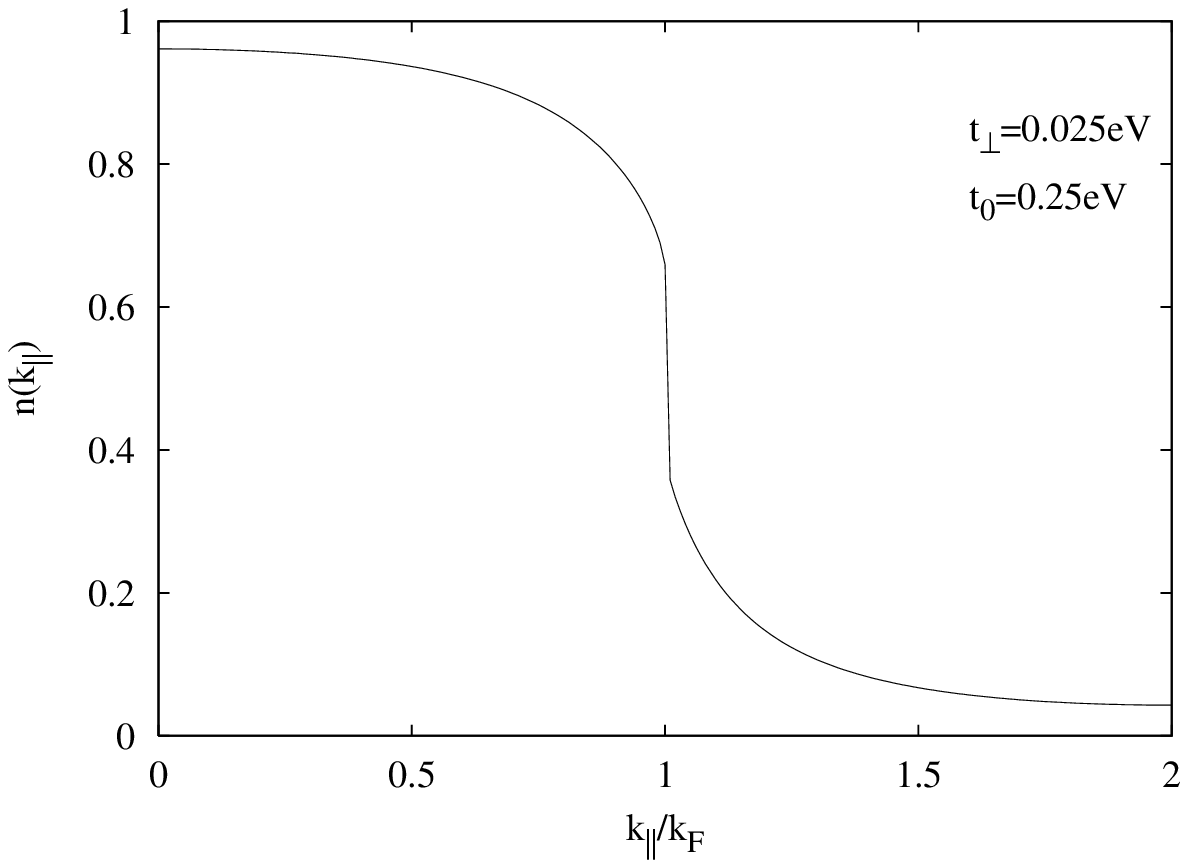}
\includegraphics{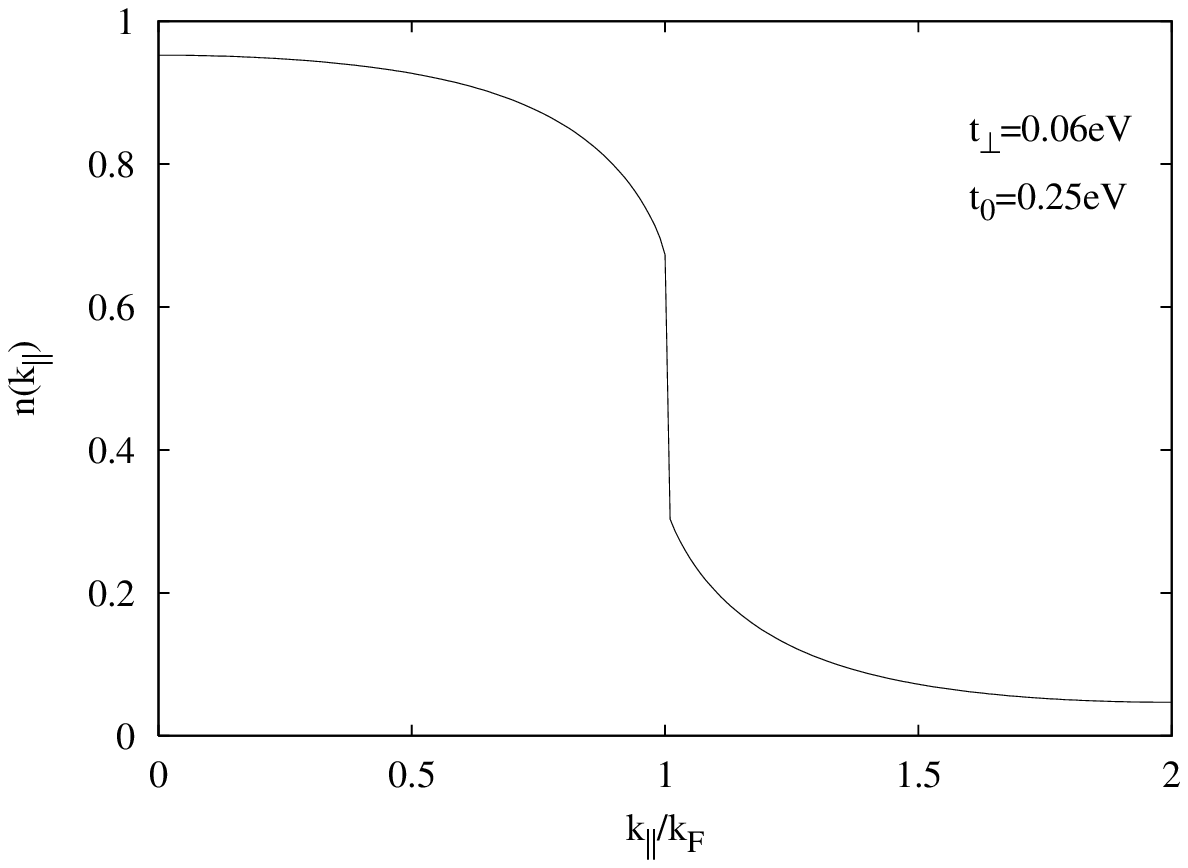} \hbox{\hspace{1.35in} (a) \hspace{3.0in} (b)}
\caption{\label{nodk}Momentum distribution function for $k_{F}=\frac{\pi}{2b}$ and $t_\perp$ equal $0.025eV$ (a) and $0.06eV$ (b)
showing the discontinuity
at $k_{F}$.}
\end{figure}

The momentum distribution function \be \label{mom1} n(k_{\parallel})=\int_{-\infty}^{\mu}A(k_{\parallel},\omega)d\omega \ee is also calculated
numerically, and shown in Fig.\ref{nodk} for $t_\perp=0.025eV$ (a) and $0.06eV$ (b). The deviation of areas below the curves (a) and (b) from
the exact number of particles is smaller than $0.1\%$, indicating the highly satisfying self-consistency of the $G_{0}W_{0}$ approximation. The
momentum distribution has a qualitative behavior of the dressed Fermi liquid. It decreases from the maximal value at $k_{\parallel}=0$ towards
the step discontinuity at the Fermi wave number $k_{\parallel}=k_{F}$. The height of this discontinuity is equal to the spectral weight
$Z(k_{F})$ of the quasi-particle $\delta$-peak at $\omega=\mu$. Fig. \ref{nodk} again shows that this height decreases as $t_\perp$ decreases.

\section{Conclusion}

The aim of the present analysis is twofold.

Firstly, we  investigate the cross-over from the specific spectral  function of one-dimensional conducting band to that of standard isotropic
three-dimensional Fermi liquid. We show that the absence of quasi-particle peaks is limited to the band with the strictly
one-dimensional flat Fermi surface. Quasi-particle peaks appear immediately with introducing a finite corrugation of Fermi surface, measured by
finite $t_\perp$ in our approach. The spectral weight of these $\delta$-peaks for $k_{\parallel}=k_F$ is given by the expression (\ref{qweight})
and shown in Fig. \ref{Zweight}. It has a non-power law dependence on the transverse bandwidth [$Z \sim - (\ln t_{\perp})^{-1}$] in the limit
$t_{\perp}\rightarrow 0$. The rest of the spectral weight is carried by the wide feature in the energy range characterized by the plasmon energy
$\Omega_{pl}$. As it is shown in Section II, this result is to a great part obtained analytically after few technical simplifications which are
well justified in the limit $t_\perp\ll t_0, \Omega_{pl}$.

Although, due to this limitation, our method of calculation cannot be extended towards pure three-dimensional regime ($t_\perp \approx t_0$),
the plausible expectation is that the quasi-particle spectral weight will increase continuously as $t_\perp$ further increases, approaching the
three-dimensional regime with quantitative properties obtained long time ago by Hedin and Lundquist \cite{Hedin,Lundq1,Lundq2}. It is worthwhile
to stress again that, as the above $Z$ \emph{vs } $t_{\perp}$ dependence illustrates, the present calculations, unlike some others (e. g.
Refs.\cite{Wen,Bourbonnais5}), are not simple power low expansion in terms of $t_\perp$, and in this respect are complementary to the
higher-dimensional bosonization approach developed in Refs. \cite{Meden1,Meden2}. The essential reason for the inadequateness of the
perturbation approach in terms of $t_\perp$, even in the limit $t_\perp \rightarrow 0$, is to be recognized in a qualitative change of the
plasmon spectrum, namely in the opening of the gap in its long-wavelength limit. This gap in turn enables the appearance of quasi-particles in
$A(\textbf{k}, \omega)$ already within the $G_0W_0$ approximation. The word of warning here concerns the applicability of the $G_0W_0$
approximation itself. Strictly, it is limited to the range of weak screened Coulomb interaction, the relevant criterion being $\Omega_{pl} <
t_0$. In some of illustrations presented here we allow for values of $\Omega_{pl}$ above this range, expecting that no qualitatively new
situation takes place in the intermediate range $\Omega_{pl} \approx t_0$. This range, as well as the range of strong long range Coulomb
interaction (even after the RPA screening taken into account) however still awaits a better understanding.

Present analysis can also provide some estimations on the possible observability of simultaneous appearance of quasi-particles and wide humps in
experiments measuring spectral properties. The energy resolution in reported photoemission measurements on Bechgaard salts varied between $10
meV$ and $30 meV$ \cite{Zwick, Zwick2,Zwick3}. Additional complication comes from indications that surface effects could have affected low
energy parts of existing ARPES data \cite{Sing}. Thus in order to observe a dispersing sharp low-energy quasi-particle with the narrow width
ranging up to $10 meV$, it will be necessary to have an increased energy resolution at low energies and an enhanced bulk sensitivity of the
ARPES spectra. We believe that such demands are achievable, particularly because our estimations suggest that the  spectral weights of
quasi-particle peaks are expected to range up to $20\%$ of the total spectral weight, and to be positioned at binding energies ranging up to the
energies of the order of $250 meV$, appearing in the coexistence with characteristic wide humps already observed at higher energies.

Among quasi-one-dimensional materials investigated in photoemission measurements the acceptor-donor chain compound TTF-TCNQ appears to be a
particularly interesting example \cite{Zwick1,Claessen}. There are various indications, like e. g. the infrared optical measurements
\cite{Tanner,Jacobsen,Basista}, that it has a soft longitudinal mode at $10 meV$ in the metallic phase. This mode was explained theoretically
within the model of the quasi-one-dimensional metal with two bands per donor and acceptor chains and the three-dimensional RPA screened
electron-electron interaction \cite{zbb}. It was shown that the appearance of such mode in the low energy range is due to the strong coupling
between the plasmon and the collective inter-band dipolar mode. As for the ARPES spectra, they show the absence of the low energy
quasi-particles and the one-dimensional dispersion of electron bands \cite{Zwick1,Claessen}. However the bandwidth values from these data are
two to four times larger than the values obtained by earlier theoretical and experimental estimates \cite{Jerome}. This signalizes that it is
necessary to include electron-electron interactions in order to improve quantitative interpretation of the data. More precisely, it remains to
investigate the influence of the elsewhere observed low energy mode on the low energy spectral properties of the quasi-one-dimensional metal
with one electron band per donor and acceptor chains within the $G_0W_0$ approximation, but with the RPA screened Coulomb electron-electron
interaction obtained for the model with two bands per chain \cite{zbb}. Taking into account the results we obtained in Ref. \cite{bbz} and in
the present work, we expect that this low energy mode is also responsible for the low energy spectral properties of TTF-TCNQ. The full analysis
of this question is under way.

\textbf{Acknowledgements.} The work is supported by project 119-1191458-1023 of Croatian Ministry of Science, Education and Sports.


\begin{references}
\bibitem{Zwick1} Zwick F {\em et al} 1998 {\em Phys. Rev. Lett.} {\bf 81} 2974
\bibitem{Claessen} Claessen R {\em et al} 2002 {\em Phys. Rev. Lett.} {\bf 88} 096402
\bibitem{Zwick} Zwick F {\em et al} 1997 {\em Phys. Rev. Lett.} {\bf 79} 3982
\bibitem{Zwick2} Zwick F {\em et al} 2000 {\em Solid State Commun.} {\bf 113}
179
\bibitem{Zwick3} Zwick F {\em et al} 2000 {\em Eur. Phys. J. B} {\bf 13} 503
\bibitem{bbz} Bona\v ci\'c Lo\v si\'c \v Z, \v Zupanovi\' c P and Bjeli\v s A 2006 {\em J. Phys.: Condens. Matter} {\bf 18} 3655
\bibitem{Hedin1} Hedin L 1965 {\em Phys. Rev.} {\bf 139} A796
\bibitem{Meden3} Meden V and Sch\"onhammer K 1992 {\em Phys. Rev. B}
{\bf 46} 15753
\bibitem{Voit2} Voit J 1993 {\em J. Phys. Condens. Matter} {\bf 5} 8305
\bibitem{Barisic} Bari\v si\'c S 1983 {\em J. Physique} {\bf 44} 185
\bibitem{Schulz} Schulz H J 1983 {\em J. Phys. C} {\bf 16} 6769
\bibitem{Botric} Botri\'c S and Bari\v si\'c S 1984 {\em J. Physique} {\bf 45} 185
\bibitem{Meden1} Kopietz P, Meden V and Sch\"onhammer K 1995 {\em Phys. Rev.
Lett.} {\bf 74} 2999
\bibitem{Meden2} Kopietz P, Meden V and Sch\"onhammer K 1997 {\em Phys. Rev. B}
{\bf 56} 7232
\bibitem{Kwak} Kwak J F 1982 {\em Phys. Rev. B} {\bf 26} 4789
\bibitem{zup} Agi\'c \v Z, \v Zupanovi\' c P and Bjeli\v s A 2004 {\em J. Physique IV} {\bf 114} 95
\bibitem{Hedin} Hedin L and Lundqvist S 1969 {\em Solid State Physics} vol 23
ed Seitz and Turnbull (Academic) p 1
\bibitem{Lundq1} Lundqvist B I 1967 {\em Phys. kondens. Materie} {\bf 6} 206
\bibitem{Lundq2} Lundqvist B I 1968 {\em Phys. kondens. Materie} {\bf 7} 117
\bibitem{Wen} Wen X G 1990 {\em Phys. Rev. B} {\bf 42} 6623
\bibitem{Bourbonnais5} Bourbonais C and Caron L G 1991 {\em Int. J. Mod. Phys. B} {\bf 5} 1033
\bibitem{Boies} Boies D, Bourbonnais C and Tremblay A-M S 1995 {\em Phys. Rev. Lett.} {\bf 74} 968
\bibitem{Clarke} Clarke D G and Strong S P 1996 {\em J. Phys. Cond. Mat.} {\bf 8} 10089
\bibitem{Tsvelik} Tsvelik A M ({\em Preprint} cond-mat/9607209)
\bibitem{Ziman} Ziman J M 1972 {\em Principles of the Theory of Solids} (Cambridge: Cambridge Univ.)
\bibitem{Artemenko} Artemenko S N and Remizov S V 2001 {\em JEPTLett.} {\bf 74} 430 (cond-mat/0109264)
\bibitem{Dessau} Dessau D S {\em et al} 1993 {\em Phys. Rev. Lett.} {\bf 71}
2781
\bibitem{Sing} Sing M {\em et al} 2003 {\em Phys. Rev. B} {\bf 67} 125402
\bibitem{Tanner} Tanner D B {\em et al} 1976 {\em Phys. Rev. B} {\bf 13} 3381
\bibitem{Jacobsen} Jacobsen C S 1979 {\em Leture Notes in Physics} vol 95
ed Bari\v si\'c et al (Springer-Verlag) p 223
\bibitem{Basista} Basista H {\em et al} 1990 {\em Phys. Rev. B} {\bf 42} 4088
\bibitem{zbb} \v Zupanovi\' c P, Bjeli\v s A and Bari\v si\'c S 1999 {\em Europhys. Lett.} {\bf 45} 188
\bibitem{Jerome} J\'erome D and Schulz H J 1982 {\em Adv. Phys.} {\bf 31} 299
\end{references}
\end{document}